\documentclass{vgtc}                          % final (conference style)
\ifpdf%                                % if we use pdflatex
  \pdfoutput=1\relax                   % create PDFs from pdfLaTeX
  \usepackage{graphicx}                % allow us to embed graphics files
  \DeclareGraphicsExtensions{.pdf,.png,.jpg,.jpeg} % for pdflatex we expect .pdf, .png, or .jpg files
\else%                                 % else we use pure latex
  \usepackage{graphicx}                % allow us to embed graphics files
  \DeclareGraphicsExtensions{.eps}     % for pure latex we expect eps files
\fi%

\graphicspath{{figures/}{pictures/}{images/}{./}} % where to search for the images

\usepackage{microtype}                 % use micro-typography (slightly more compact, better to read)
\PassOptionsToPackage{warn}{textcomp}  % to address font issues with \textrightarrow
\usepackage{textcomp}                  % use better special symbols
\usepackage{mathptmx}                  % use matching math font
\usepackage{times}                     % we use Times as the main font
\usepackage{cite}                      % needed to automatically sort the references
\usepackage{tabu}                      % only used for the table example
\usepackage{booktabs}                  % only used for the table example
\usepackage{dblfloatfix}
\usepackage{xcolor}
\usepackage[normalem]{ulem}
\usepackage{tabularx}  

\onlineid{1484}

\vgtccategory{Research}

\vgtcinsertpkg

\title{DocuBits: VR Document Decomposition for Procedural Task Completion}

\author{Geonsun Lee\thanks{e-mail: gsunlee@umd.edu}\\ %
        \scriptsize University of Maryland %
\and Jennifer Healey\thanks{e-mail: jehealey@adobe.com}\\ %
     \scriptsize Adobe Research%
\and Dinesh Manocha\thanks{e-mail: dmanocha@umd.edu}\\ %
     \scriptsize University of Maryland}

\teaser{
  \includegraphics[width=\textwidth]{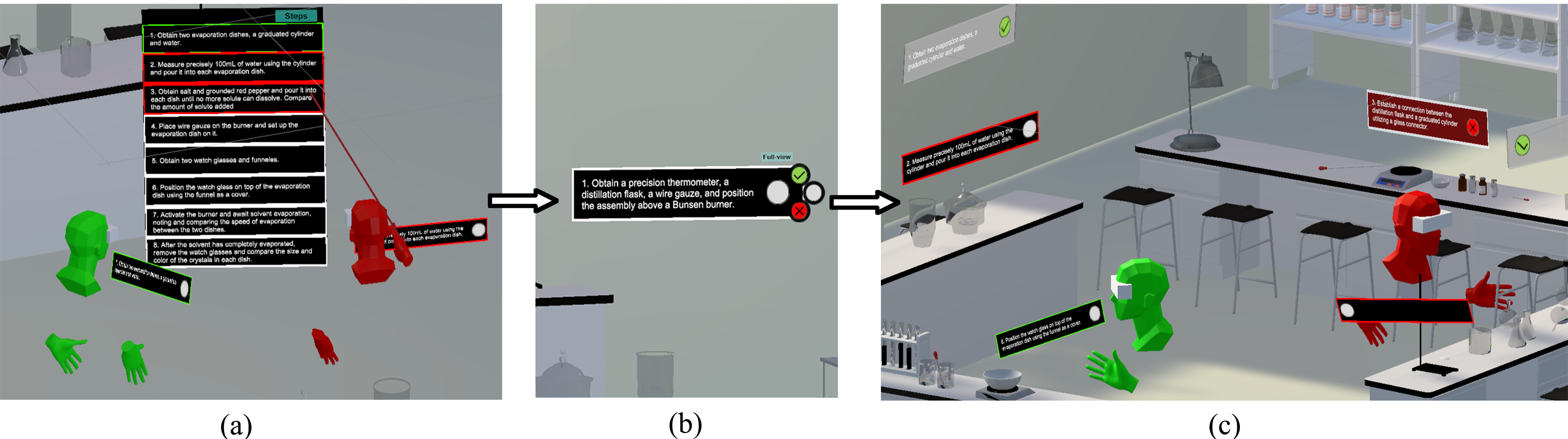}%[width=1.5in]
  \vspace{-1em}
  \caption{DocuBits: Interactive document fragments strategically anchored to pertinent task locations, minimizing attention shifts between reading and task execution, and fostering collaborative learning. In a multi-user context, participants can (a) fragmentize and select specific sections of text instructions for execution, (b) interact with the DocuBit to monitor progress, and (c) position these fragments within the virtual environment for seamless sharing among all users.}
}

\abstract{Reading monolithic instructional documents in VR is often challenging, especially when tasks are collaborative.  Here we present DocuBits, a novel method for transforming
monolithic documents into small, interactive instructional elements.
Our approach allows users to:(i) create instructional elements
(ii) position them within VR
and (iii) use them to 
monitor and share progress in a multi-user VR learning environment. We describe our  design methodology
as well as two user studies evaluating how both individual users and pairs of users interact with DocuBits compared to monolithic documents while performing a chemistry lab task. Our analysis shows that, for both studies, DocuBits had substantially higher usability, while decreasing perceived workload ($p < 0.001$). Our collaborative study showed that participants perceived higher social presence, collaborator awareness as well as immersion and presence ($p < 0.001$). We discuss our insights for using text-based instructions to support enhanced collaboration in VR environments.}

\CCScatlist{
  \CCScatTwelve{Human-centered computing}{Human computer interaction (HCI)}{Interaction paradigms}{Virtual reality};
  \CCScatTwelve{Human-centered computing}{Human computer interaction (HCI}{Interaction techniques}{};
  \CCScatTwelve{Human-centered computing}{Human computer interaction (HCI}{Interaction paradigms}{Collaborative interaction};
}

\begin{document}


%\firstsection{Introduction}

\maketitle

%% \section{Introduction} %for journal use above \firstsection{..} instead

\section{Introduction}
%demonstration is common in VR
%demonstration benefits (task at hand), requires sequential consumption, not glanceable, hard to remember and refresh without replaying it
%standardized instruction given through texts. people should go through an interpretation of the information.
% thinking of the game world and not necessarily about the information itself.
%demonstration is unique to a certain demonstrator there is now ay to validate the instruction given through it.
% also allows you to author and standardize seamlessly, and people who doesn't know how to play in VR can still (author burden is a lot easier with text)
Virtual Reality (VR) can provide a safe place for people to learn how to perform tasks that would be difficult, dangerous, or expensive to practice in the real world, e.g., flight simulation, surgery, cooking, or laboratory experiments. The most common way that people learn to perform a task is with the help of a set of written instructions.  Even when augmented by live teacher demonstration, written instructions provide a common standard for how a process should be completed. Many institutions seek a standardized way of training workers and many school systems have a common core of learning to ensure similar instruction across schools. Our goal is to provide a method to retain all the benefits of this kind of standardized instruction while completing it and extending instruction to virtual environments.  

Currently, most virtual reality instruction is customized to the application with specialized follow-along demonstrations incorporated into the virtual environment~\cite{patel2006effects}.  This kind of instruction is generally preferred because it is simple to follow and takes advantage of the richness of the virtual world.
%However, there are two major drawbacks to this approach.

We note, however, that there are two possible drawbacks to this approach.
The first is that some processes already have set of text instructions used in physical lab training and that these might be periodically updated.  With DocuBits, the current, approved text can be directly brought into the VR environment without having to re-make a demonstration video.
%text
%Assuming that some students will be taught in live settings and some will be taught in VR,  the difficulty of standardizing the demonstration.  
%The first is that it is difficult to standardize this type of instruction.  It is specific to the particular version of the virtual reality application in which it exists.  Every new application must be evaluated to ensure that it conforms to a particular standard, and as technology advances and applications become obsolete, this type of instruction becomes inaccessible.  
The second is that demonstration might inspire mimicry rather than challenging
%drawback is that because demonstration instruction is so easy to mimic, 
%it does not challenge 
the learner to reason about the instruction in an abstract way. 
%They can come away with less understanding about the task than if they had performed the task from written instructions.  
We believe that the benefit of text-based instructions is that they are identical across physical and virtual environments, they are abstract, and they remain human-readable indefinitely.

%The challenge with written instructions is that reading long text is often difficult in 

Reading text in VR also presents its own set of challenges, especially when text is longer and when tasks are shared.  These include: reading itself, task switching, tracking progress and communicating progress.
%VR
~\cite{kojic2020user, kobayashi2021examination, gabel2023immersive}. 
%In addition, performing procedural tasks in VR  often poses the problem of how best to read the 
DocuBits addresses the challenge of reading longer text by simply breaking up the text into smaller units.  These smaller text units are then placed withing the interactive DocuBit and can be placed strategically within the environment.  With smaller bits of text, the reader can read once and simply glance back to refresh their memory about the current task step, allowing them to transition more freely between reading instructions and doing the work.   
%Often instructional documents are long and require users to switch contexts between reading and doing.  
%In complex tasks, 
The interactive aspect of DocuBits helps readers keep track of their work with different display modes for unattempted instructions, completed tasks and failed or blocked tasks.  These different displays also simplify communication between users working on shared tasks.

%This especially degrades task performance involving multiple users working together because they need to verbally communicate to see where the other user is reading. 
%During a complex task, 
%It is often difficult to remember which steps have already been completed, which steps have yet to be tried, and which steps have been attempted but could not be completed successfully because of some blocking issue. 

We believe that this interactive display aspect is key to the value of DocuBits and that it will allow multiple participants to share in both reading and doing tasks equally as well as provide an overall ``birds-eye" view of progress to a third party, such as a teacher, who might want to oversee the users experience.  This is an improvement to some current lab situations where the problem of task switching is solved by one person taking a "reader" role and another taking a ``doer" role (resulting in an uneven experience) and where teachers must go around to each group to assess progress instead of being able to get a quick overall view.
%With some current lab configurations, configurations overseeing all participants 
%This often leads to users assigning one to read the instruction for the other, resulting in unequal experience in task training. 

%A better solution is to break such documents up into smaller units of text and allow the relevant text to be placed next to where the task needs to be performed.  
%DocuBits addressed both the document length issue and the progress tracking issue by creating short, interactive instructional elements that reflect task progress to all users.
%In this way, the user can simply glance at the relevant text and quickly look back at the task.  
%Moreover, users can visually see how the document is distributed which allows them to be involved in the tasks equally. Another problem that arises is how to communicate the current state of a task.  During a complex task, it is often difficult to remember which steps have already been completed, which steps have yet to be tried, and which steps have been attempted but could not be completed successfully because of some blocking issue. 

%Written instructions provide a practice is that  that people learn to perform tasks is by following written instructions, but reading monolithic documents is difficult in VR.  
\noindent {\bf Main Results:} We address the challenge of following procedural documents in VR with DocuBits, a novel method for transforming monolithic instructional documents into text fragments and embedding those text fragments into portable interactive virtual objects.
%Our invention, DocuBits, addresses both these problems with a system and method for deconstructing monolithic text documents into
These virtual text elements (text segments) can be moved and placed next to virtual task spaces that comprise both active elements (multi-state checkboxes), which specifically convey the state of the task, and state-based behaviors (floating up, fading away, bouncing up and down), which communicate the current state of a task from a distance (especially beneficial for multi-user collaboration). Not only does the user have a sense of action history, but collaborators can proactively support each other as the progress is visually shared. 

In this paper, we present our methodology for the design and development of DocuBits, a technical description of our system, and two user evaluations, one with an individual user and one with pairs of users who interact both with the DocuBits and each other.  We use a fictional chemistry lab experiment as a testbed for our evaluations while envisioning the broad applicability of our method to any procedural task suitable for VR instruction. Our evaluations show that DocuBits both improves the immersion and lowers the cognitive load when compared to a baseline of non-segmented, monolithic documents of instructions. In the paired-users evaluation, participants report higher perceived levels of co-presence and collaboration. This demonstrates  DocuBits' potential to improve the effectiveness of VR training for procedural tasks and enable multi-user collaboration. In summary, our contributions include:
\vspace{-0.1em}
\begin{itemize}
    \itemsep-0.1em
    \item Uncovering insights from real-life instructors on the challenges associated with text instructions in educational settings, thereby informing design considerations for VR interfaces.
    
    \item Presenting DocuBits, an innovative document interface that allows fragmentation, interaction, and anchoring within the VR space, enhancing the effectiveness of reading-while-doing.
    
    \item Introducing four user interactions: (i) Doc to Bits; (ii) Tag Along and Stick; (iii) Progress animation; and (iv) Assignment, which are combined for the overall interactive approach.
    
    \item Conducting a comprehensive user study on DocuBits, encompassing both single-user and paired-user scenarios.
    
    \item Demonstrating the significant impact of DocuBits on user experience, particularly in collaborative learning scenarios, and extracting valuable insights on user behavior with text instructions from the study.
\end{itemize}

\section{Related Works}
%\subsection{VR for Education and Training}
\subsection{VR for Training and Education}
Virtual reality has often been used to train people to perform tasks in cases where tasks are dangerous, difficult, or expensive to replicate.  Examples of this include high-impact activities such as surgical work, firefighting~\cite{tate1997virtual}, first responder training~\cite{stansfield1999biosimmer}, aircraft and spacecraft control~\cite{loftin1995training}, and construction work~\cite{adami2021effectiveness}. Additionally, VR has proven effective in more routine yet skill-intensive tasks such as assembly and maintenance. For instance, VR training systems have been developed for automotive assembly tasks~\cite{garcia2009simulation,becker2011using}, offering a safe and cost-effective method for training workers in complex assembly processes. In the field of maintenance, VR has been used to train technicians in the maintenance of industrial equipment, reducing the risks and costs associated with hands-on training~\cite{gavish2015evaluating,bailey2017using}. This flexibility of virtual reality also extends to other common tasks such as cooking~\cite{gao2019vrkitchen}, Tai Chi~\cite{chua2003training}, and running a chemistry laboratory~\cite{georgiou2007virtual}, where time and materials make repeated practice costly.

VR's on-demand instruction allows users to practice tasks flexibly, free from traditional classroom constraints. This adaptability supports individual learning needs and pacing~\cite{patel2006effects}. Its key advantage lies in reducing social pressures for better skill focus and lowering resource use compared to physical training environments. This is particularly beneficial in scenarios with limited face-to-face interaction~\cite{trondsen1997learning}.

For educational applications of VR, Dieterle and Clark created River City, a virtual environment for fostering scientific inquiry and skills in middle school science classes~\cite{dieterle2009multi}. Stokes et al. investigated attention distraction in children with ADHD within a simulated VR classroom, employing eye-tracking measures to overcome traditional laboratory method limitations~\cite{stokes2022measuring}. Parsons et al. utilized a VR classroom setting, investigating its potential for controlled performance assessments in an ecologically valid context, highlighting the impact of distraction stimuli when contrasted with standard neuropsychological methods~\cite{parsons2007controlled}.

%The flexibility of virtual reality enables on-demand instruction, permitting users to practice tasks as frequently as necessary, independent of the constraints imposed by a traditional classroom environment. This adaptability tailors learning to the user's individual schedule, goals, and pace~\cite{patel2006effects}. 
%The distinct advantage of learning on demand lies in its ability to alleviate social pressure, allowing users to focus on skill acquisition. 
%The distinct advantage of learning on demand lies not only in its ability to alleviate social pressures, thus enabling users to concentrate more effectively on skill acquisition, but also in its potential to significantly reduce the resources required compared to organizing learning or training in a physical environment.
%This is especially valuable in scenarios where face-to-face human interaction may be impractical or limited~\cite{trondsen1997learning}.

%The adaptability of VR allows for tailored, on-demand instruction, ideal for complex tasks in assembly and maintenance. It's effective in a range of scenarios, from everyday tasks to specialized activities where traditional training is impractical~\cite{chua2003training, trondsen1997learning}. VR's advantage extends beyond enhancing skill acquisition; it also offers a resource-efficient training alternative, reducing the need for physical resources and environments.
%Most other VR instructions are either VR/MR-based or video-based \cite{cai2014case,chidambaram2021processar}.

%\subsection{Collaborative Learning in VR/AR}
\subsection{Collaborative Work in VR/AR}
%There are previous works proving 
Prior work has proven the positive impact of VR and AR in collaborative and distance education settings~\cite{10098484}. Simeone et al.~\cite{simeone2019live} compared a system-based teaching system with a single-user and a two-user version; the latter had one user acting as a teacher. The study results revealed that the two-user version scored higher in overall preference and clarity compared to the single-user version, which used animation sequences for teaching.

Radu et al.~\cite{radu2021unequal} found that in a robot peer-programming task, augmenting informatic visualizations through AR headsets led to higher team learning gains compared to learning without visualizations. Some works researched collaborative learning scenarios where users use different devices and have different roles assigned. For example, a table is given to a user who has the guiding role within an educational game, similar to that of a teacher~\cite{thompson2018rules, uz2020exploring, lee2020rolevr}. Drey et al.~\cite{drey2022towards} developed VR pair-learning systems and demonstrated that it is preferable to have a symmetric system, with both participants wearing a VR HMD, than an asymmetric system, with one student using a tablet and the other a VR HMD, in terms of immersion, presence, and cognitive load. However, these works neglect the role of text presentation in their scenarios. Additionally, they mostly consider a teacher-student relationship rather than a peer-to-peer learning scenario which is what DocuBits focuses on supporting.

Schott et al.~\cite{schott2021vr} built a VR/AR multi-user environment for liver anatomy education, considering how certain types of information should be presented in the environment, such as visualizing 3D models in a shelf-like object and presenting text and image data on a shared board. Jin and Lie et al.~\cite{jin2023collaborative} investigated collaborative learning on VR video viewing systems, comparing viewing modes based on whether users have individual control or shared control over video sync. Results showed that shared VR video modes increase collaboration and social presence. Loki~\cite{thoravi2019loki} is a system that supports remote instructions by capturing video, audio, and spatial information for a mixed-reality presentation of the learner and the listener. It allows learners to record their performance so that a remote instructor can review and annotate them in AR and VR. While these works have concentrated on demonstration-led learning, DocuBits shifts the focus to the often-overlooked importance of text-based instruction, enhancing learning where following the documentation is key.

\subsection{Text Presentations and Interfaces}
Investigating legibility in text presentation on electronic devices has been a longstanding concern, with Dillon et al.~\cite{dillon1990effects} addressing its importance. In virtual environments, text presentation encounters unique challenges due to resolution constraints and the introduction of a third dimension.  Chen et al.~\cite{chen2004testbed} evaluated different combinations of VR navigation techniques and text layout techniques in Information-Rich Virtual Environments. Dittrich et al.~\cite{dittrich2013legibility} proposed guidelines for text visualizations in 3D virtual spaces, emphasizing the need for larger text sizes than those on 2D displays. Jankowski et al.~\cite{jankowski2010integrating} explored the integration of text with video and 3D graphics, revealing preferences for negative text presentations. Dingler et al.~\cite{dingler2018vr} investigated optimal parameters for VR text presentation, including text size, convergence, and color. Recent efforts include studies on text presentations on 3D objects with various surfaces~\cite{wei2020reading}, revealing the ease of reading text wrapped around a 3D object with a single axis.

While reading performance on electronic devices has been extensively compared~\cite{dillon1992reading, gould1987reading, chen2014comparison, connell2012effects, delgado2018don}, Rau et al.~\cite{rau2018speed} evaluated reading speed and accuracy in VR and AR, finding differences compared to traditional computer screens. Some works investigated optimizing interfaces for reading long texts in VR. 
Kojic et al.~\cite{kojic2020user} tested different values for text parameters such as font size, distance, type of HMDs used, etc., for short, medium, and long text samples. Kobayashi et al.~\cite{kobayashi2021examination} explored text parameters for reading long texts and proposed view settings for better readability and less fatigue in VR. Gabel et al.~\cite{gabel2023immersive} compared four text panel UIs to present long texts, finding no significant differences in reading performance. Lee et al.~\cite{lee2022vrdoc} developed a set of gaze-based interactions to improve the user's reading experience in VR, showing enhanced effectiveness and less perceived workload.

However, these works only consider the sole activity of reading and not how users interact with texts when they are performing tasks based on them. DocuBits proposes how we can support the user in such experiences and even multi-user scenarios for this.

\section{Initial Concept and Design}
%Our initial concept involved fragmenting documents into smaller units and strategically placing them near corresponding tasks to enhance the accessibility and standardization of text in virtual reality (VR). To identify additional desirable attributes for these text fragments, we engaged a group of experienced educators to explore the specific requirements of learners when following instructional procedures.
Our concept focuses on fragmenting documents into smaller units, strategically placed in VR to enhance text accessibility. To refine this concept, we consulted experienced educators for insights into learners' needs in instructional settings.

\begin{figure*}
  \includegraphics[width=\textwidth]{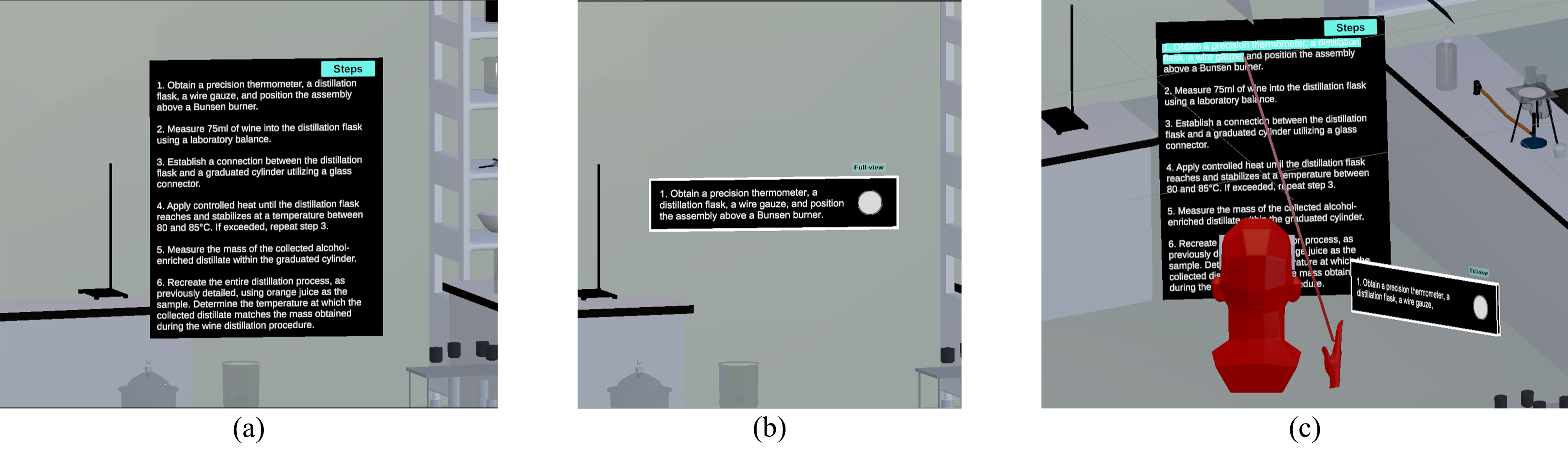}%sampleteaser
  \caption{Our method for creating Doc to Bits. (a) The initial monolithic document appears, and users can either (b) create DocuBits with pre-defined numbered steps or (c) customize DocuBits by highlighting and releasing words.}
  \label{fig:single-user}
  \vspace{-2em}
\end{figure*}

\subsection{Needs-Finding Study}
\label{sec:needs-finding}
To develop a set of design principles for DocuBits beyond the idea of segmented text we recruited a focus group of 7 experienced educators in three different areas. The areas of expertise include physics (E1), biochemistry (E2, E3), computer science (E4, E5), jewelry craft (E6), and landscape design (E7). All the participants had 3 to 6 years of experience in teaching students and trainees and often followed instructional documents such as class projects and laboratories. All participants spoke the same language, and the focus group interview was conducted in two separate sessions to accommodate their availability. 

During the interviews, we focused on several key areas: methods of distributing instructional documents to students, observations of student interactions and collaboration with these documents, frequency of students referring to instructions during tasks, and educators' involvement in student help requests. We also inquired about students' collaborative dynamics, such as role division in group tasks, and identified the main challenges students encountered while following procedural instructions, including understanding the written content and managing collaborative efforts.

Key takeaways included the educators' use of both physical and digital documents and the dynamics of student interactions when using reading materials in group tasks. Common queries from students were about locating necessary objects and the practical difficulties in handling documents in lab settings, such as exposure to fluids and managing them with gloves or protective eyewear.

We present the results of the interview study in terms of the benefits and drawbacks of utilizing documents in instructing students in the following paragraphs. %The interviews highlighted benefits like aiding self-paced learning and facilitating content augmentation and reformatting. However, drawbacks such as cognitive burdens in task switching, uneven role distribution in group tasks, and difficulty in tracking student responsibility were noted. We provide further details for each of these findings in the following paragraphs.

%%jen - training in their own pace - this paragaph is talking about the students being able to go at their own pace.  Tecnically students are the trainees not the trainers so 
%\paragraph{Training in their own pace}: Depending on the interviewee's teaching style, the document was either distributed in a physically-printed form or a digital form. In the case of a digital form, the instructions were distributed in a pdf file where all of the steps are listed on a single page or divided into steps with supporting figures. All participants agreed they distribute instructional documents instead of demonstrating each step when there are a large group of students which makes it difficult to follow each student's progress. P2, P3, and P6 pointed out that they usually present a demonstration of the entire process to the class, and then allow students to take their own pace afterward by following the steps in the document. 

%a paragraph that goes through the overview of the benefits and drawbacks before going into the details. We present the result of the formative study in terms of benefits and draw backs of utilizing documents in instructing students. Benefits include...

\subsection{Benefits of Utilizing Text Instructions}
%\begin{itemize}
%    \item Self-paced learning: Instructional documents allow students to reference and follow the steps of a process at their own pace, giving them control over the pace of  learning.
    %speed and understanding.
%    \item Progress tracking: Students can use documents to keep track of their progress, mark completed steps, add annotations, and make notes. 
%    \item Content augmentation and reformatting: Students often add annotation and make notes on paper documents, in digital documents students to edit and create their own materials.
%\end{itemize}
\textbf{Self-paced learning} All educators agreed that they distributed instructional documents instead of demonstrating each step in a process because a document provides a concrete reference for the steps. 
Demonstrations often demand continuous viewing up to a specific time step for comprehension and necessitate re-watching from the beginning if details are forgotten, making them more time-consuming compared to text-based instructions.
%Demonstrations are time-consuming and sometimes difficult to follow and remember precisely. 
%In addition, when there is a large group of students it is difficult to know to what extent each of them is understanding the demonstration. 
E2, E3, and E6 pointed out that they usually do a demonstration of the process to the class and then allow students to go at their own pace afterward by referencing and following the steps in the document.

\textbf{Content augmentation and reformatting} To mitigate situations where the students lose track of their own progress or need their own interpretation to help them perform the task, students augment the document with actions such as adding drawings and crossing out steps that are finished or highlighting parts that need to be revisited. When accessing digital documents, there are cases in which students go further by editing and creating their own documents so the task is decomposed or integrated in a way that is more convenient for their own workflow.

Overall, although the interviewers unanimously recognized the advantages of employing distributed instructional documents for teaching students, there is a notable dearth of tools available to facilitate the efficient utilization of such documents by students. This issue becomes even more pronounced in collaborative learning scenarios.

\subsection{Drawbacks in Utilizing Text Instructions}
\textbf{Losing track of progress} The majority of the educators (E2, E3, E4, E5) noted that students frequently get ``lost'' in their progress when following documents, particularly when switching between tasks, changing positions, or asking questions. E5 remarked, ``When a student had a question to ask, there was a time gap until I came over, and then the student had to skim through the document again to find the point they wanted to refer to." E2 added, ``This happens even when they are following the document themselves, as I once observed a student redoing steps to trace back their mistakes."

\textbf{Unequal role distribution of reading and doing} In collaborative experiments, E2 and E3 observed scenarios where one student read out instructions for another performing the task. E1, E4, E5, and E6 noted in larger group projects, often one student would take charge of tracking progress, leading to an unequal distribution of understanding among the team. This was a concern for five of the interviewees as it could lead to a skewed comprehension of the task across the group.

\textbf{Credit Attributing} Instructors who incorporated physical movement within instructional procedures (E2, E3, E6, E7) highlighted difficulties in identifying which student was responsible for specific tasks, especially when using online documents. This often made it challenging for instructors to offer timely assistance and for peers to support each other effectively in the learning process.

In conclusion, while instructional documents offer advantages such as personalized learning and progress tracking, there are notable challenges in managing student understanding, collaboration, and prompt assistance in collaborative learning scenarios. We use these observations to design our interface in Section~\ref{sec:method}.

\subsection{Design Considerations for Docubits}
Based on the group interview insights, we identified key design principles for DocuBits, focusing on the need for collaborative tools in instructional document use.

\textbf{Dynamically fragment documents:} 
Considering diverse learning paces, it is suggested to design DocuBits to allow students to dynamically fragment documents into smaller segments like numbered steps or sentences. This flexibility is vital for catering to both individual preferences and group collaboration needs, allowing for customized information flow.

\textbf{Spatial control:} 
Given the varying spatial dynamics in learning environments, DocuBits should enable users to control the positioning of documents effectively. This feature would ensure that whether students are stationary or moving, they have continuous and easy access to instructional content.

\textbf{Visibly share progress:} 
Incorporating interactive elements within documents, such as checklists and drawings, is recommended for efficient progress tracking. The implementation should include visible indicators of task status and progress, which would enhance learning experiences by minimizing the need for verbal communication in both individual and group settings.

\section{DocuBits: Interactive Spatialized Text Instructions}
\label{sec:method}

In accordance with our design considerations, we developed and implemented DocuBits, an interactive solution for users to effectively engage with instructional documents in VR training scenarios. Our goal was to create a versatile framework that worked well for both individual and group-based learning contexts.  We wanted to implement and test several experiences that we believed would address the needs uncovered in our formative study. To this end, we enabled four key experiences: (i) Doc to Bits a method for fragmenting a procedural document into cohesive units of texts representing tasks; (ii) Tag Along and Stick;  a method for easily placing those bits in the environment; (iii) Progress Animation, a method for visibly expressing task progress; and (iv) Assignment, a method for associating bits to a particular user.  

Our approach has been developed using the Unity game engine with Photon~\cite{Photon} for multi-player networking.  To deliver an immersive VR experience, we use two Oculus Quest 2 head-mounted displays (HMDs) per user. The virtual environment has been meticulously set up to resemble a chemistry lab, featuring scattered equipment within the spatial domain.  

\subsection{Doc to Bits}
We enable two experiences for users to convert monolithic documents into bits, one that automatically segments the document by numbered steps and one that allows custom separation by highlighting. Upon entering the VR environment, the instructional documents initially appear as a regular ``monolithic" document (Figure.~\ref{fig:single-user}-(a)). In our use cases, all documents had numbered steps, but we believe many procedural documents with other kinds of separators, dashes, dots, lines, etc. can be converted into documents with numbered steps in a straightforward manner. Using the first approach, the user simply clicks on a button labeled ``steps'' and the monolithic document automatically transforms into a stack of DocuBits, one for each numbered instruction (Figure.~\ref{fig:single-user}-(b)). The second approach is custom highlighting.  This approach is useful when users want more granular control over what parts of the document should be individual DocuBits (Figure.~\ref{fig:single-user}-(c)).  For example, some documents may contain long steps that might be better as two or three DocuBits, while some steps might include more text than is necessary for a DocuBit. Highlighting offers this fine-grain editing control while still only requiring the user to swipe over the desired text using a press-and-drag action. Upon releasing the controller button, the highlighted segment is recognized as a single DocuBit.  All created DocuBits are displayed as stacks that remain fixed to the user's avatar position, facilitating ease of movement within the VR environment.
Note that users can switch to a ``full view'' of the document as they desire.  Figure~\ref{fig:single-user} shows the method and the user's views.
%\begin{itemize}
%    \item Two different methods:
%    \item Select pre-defined blocks based on numbered steps within the document.
%    \item Create arbitrary segments by selecting texts using the controller.
    
%\end{itemize}
\subsection{Tag-Along-and-Stick}
Once a stack of DocuBits is created, users can sort the stack and then take any individual Docubit, stack, or sub-stack with them.  The DocuBits ``tag along" behind the person's avatar as they navigate the space.  The user is then free to grab a DocuBit and place it in the environment where they feel the instructions would be most helpful.  Once placed, the DocuBit positions can be saved with the environment.  This would enable, for example,  a teacher to set up the DocuBits for students in advance of a lesson to save time. It also enables students to save progress if procedures need to be interrupted and resumed later. %One option for authors is to place the DocuBits in the environment and then save the placed locations as part of the environment. The DocuBits will then be in place when the next user starts the environment.

In DocuBits, users can switch between different DocuBits and affix them within the virtual environment. When a DocuBit is placed or ``stuck" in a specific location, it remains stationary. However, if the corresponding task associated with the DocuBit is still in progress (not marked as finished), an additional DocuBit clone is generated and follows the user when the initially placed DocuBit is no longer within the user's view frustum (Figure.~\ref{fig:progress}-(c)).  Furthermore, users can revisit previously encountered DocuBits by flipping through them. These past Docubits are visually distinguished by a gray tint, indicating their completed status.

%\begin{itemize}
%    \item Users may flip between Docubits and `stick' them within the virtual environment.
%    \item Even when the Docubit is `stick', if the task is still in progress (not marked as finished), the Docubit will clone itself and follow the user when the placed Docubit is out of sight from the user's view frustum.
%    \item Users can still view their past Docubit by flipping through which is tinted gray.
%\end{itemize}
%\subsection{Sharing Progress}
\subsection{Progress Animation}
%\begin{itemize}
%    \item Work-in-progress/ Finished/ Unfinished
%    \item Floating Behavior
%\end{itemize}
\begin{figure}
  \includegraphics[width=0.5\textwidth]{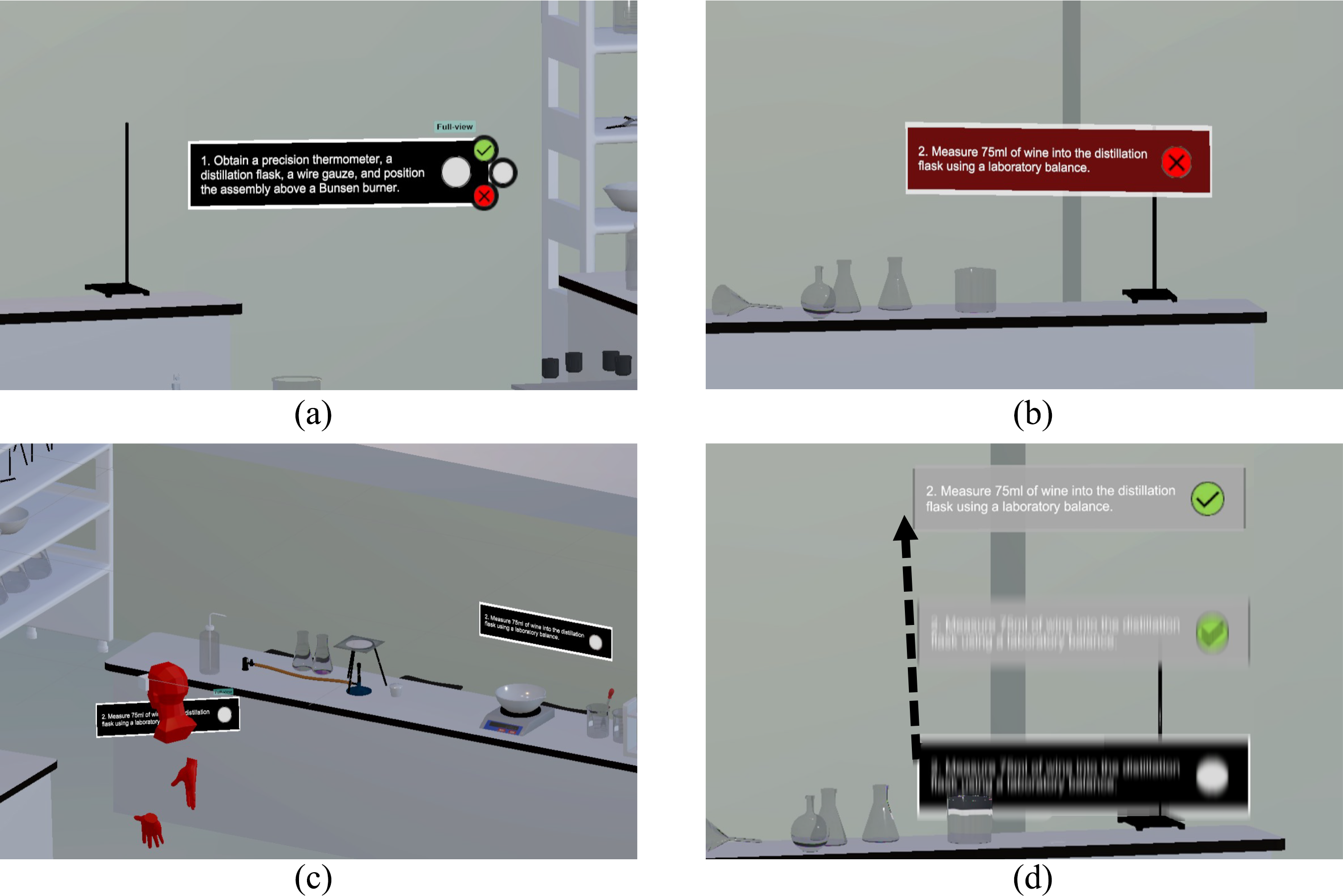}%sampleteaser
  \vspace{-0.5em}
  \caption{DocuBits are interactive text elements that maintain and display the status: (a) Users can select a status such as: (b) task attempted but not successfully completed, (c) task in progress, or (d) task successfully completed. When a DocuBit is still in progress, a clone of a DocuBit will tag along with the user as in (c). Depicted in (d) is the animation of a complete DocuBit floating upwards.}
  \label{fig:progress}
  \vspace{-2em}
\end{figure}

Each DocuBit also has a status indicator and associated animation behaviors that show if a task is not attempted, attempted but not completed, or successfully completed (Figure.~\ref{fig:progress}-(a)). Steps that have not yet been attempted have a clearly visible white color and are placed near a task. Upon successful completion, a green light appears in the status indicator circle, the body of the object turns gray, and the object floats up (Figure.~\ref{fig:progress}-(d)).  If the person cannot complete the task (for example, the user has a question about the instructions or cannot determine how the instructions relate to the virtual world) the user can toggle the status indicator to red and the DocuBit will present with a red indicator and fully visible text (Figure.~\ref{fig:progress}-(b)). These interactive behaviors are designed to capture users' attention intuitively regarding users' overall progress. While the completion of an action can be automatically detected depending on the task scenario, in our implementation we allowed users to manually select the status of the DocuBit. Such indications are valuable for both multi-participant task teams and for teachers who are overseeing the progress of multiple students in the virtual classroom.

\subsection{Assignment}

\begin{figure}
  \includegraphics[width=0.5\textwidth]{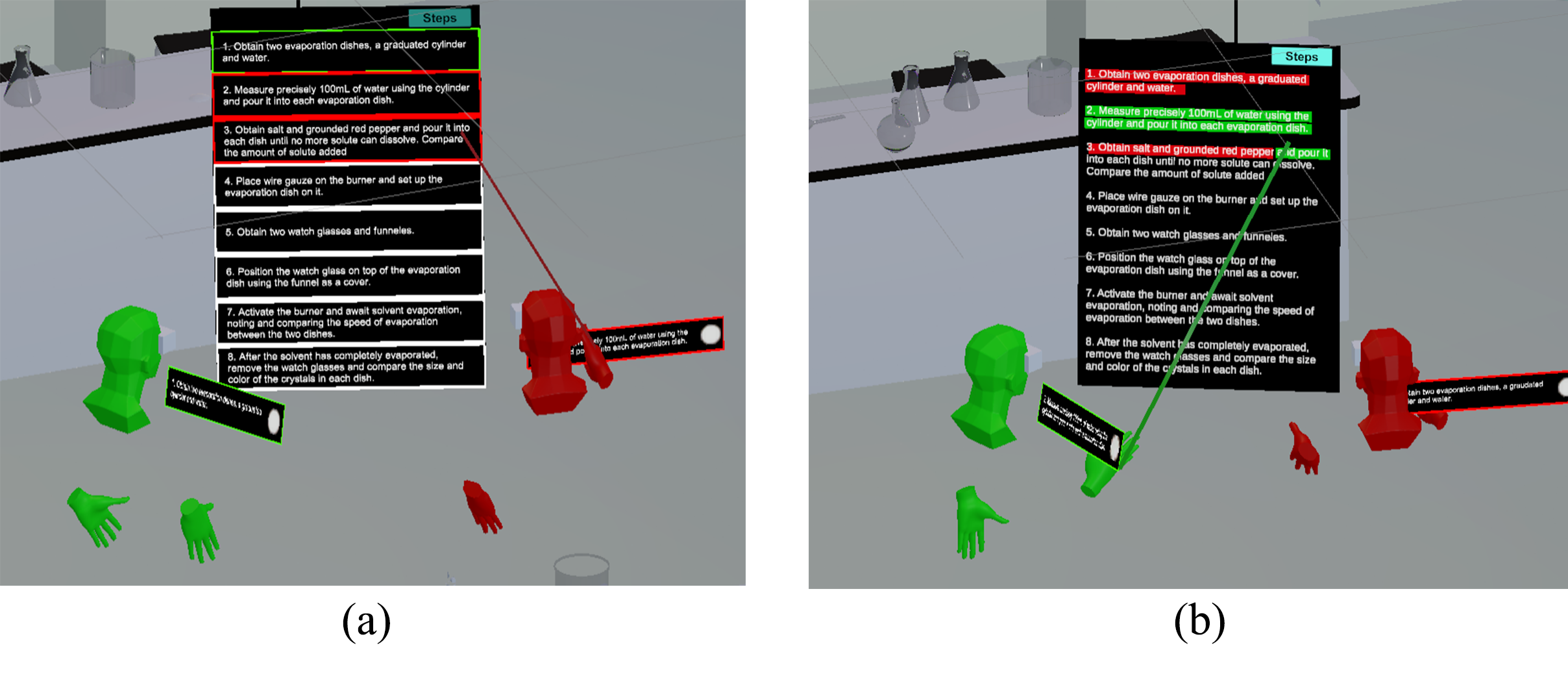}%sampleteaser
  \vspace{-1em}
  \caption{Users can allocate ownership of the DocuBits between multiple users by either (a) selecting pre-defined steps or (b) highlighting words.}
  \label{fig:ownership}
  \vspace{-1em}
\end{figure}

In collaborative settings, when multiple users are performing a task together,
%engaged,
the assignment of DocuBits to individual users becomes crucial. To denote ownership, distinct color codes are employed, such as red for User A and green for User B. The ownership allocation occurs during the initial fragmentation of the monolithic document.
%For instance, if 
If users opt to fragment the document automatically based on pre-defined steps (Figure.~\ref{fig:ownership}-(a)), one participant 
%% jen added this part about number of users in the future wishful hopeful tense "would" (in some imagined future) unless this is actually implemented
would need to specify how many people will be participating. Then the selected blocks of content will be 
%%more adds
divided into task-cohesive stacks in front of each user.  Currently, task-cohesive steps need to be specifically identified by the document author, but ideally, these could be logically inferred in some future implementation.  If users choose to create DocuBits by highlighting specific words (Figure.~\ref{fig:ownership}-(b)), each DocuBit is assigned to the participant who created it after the user selects and releases the trigger button to highlight a sentence.  
%automatically becomes that user's possession.

\begin{figure}
  \includegraphics[width=0.5\textwidth]{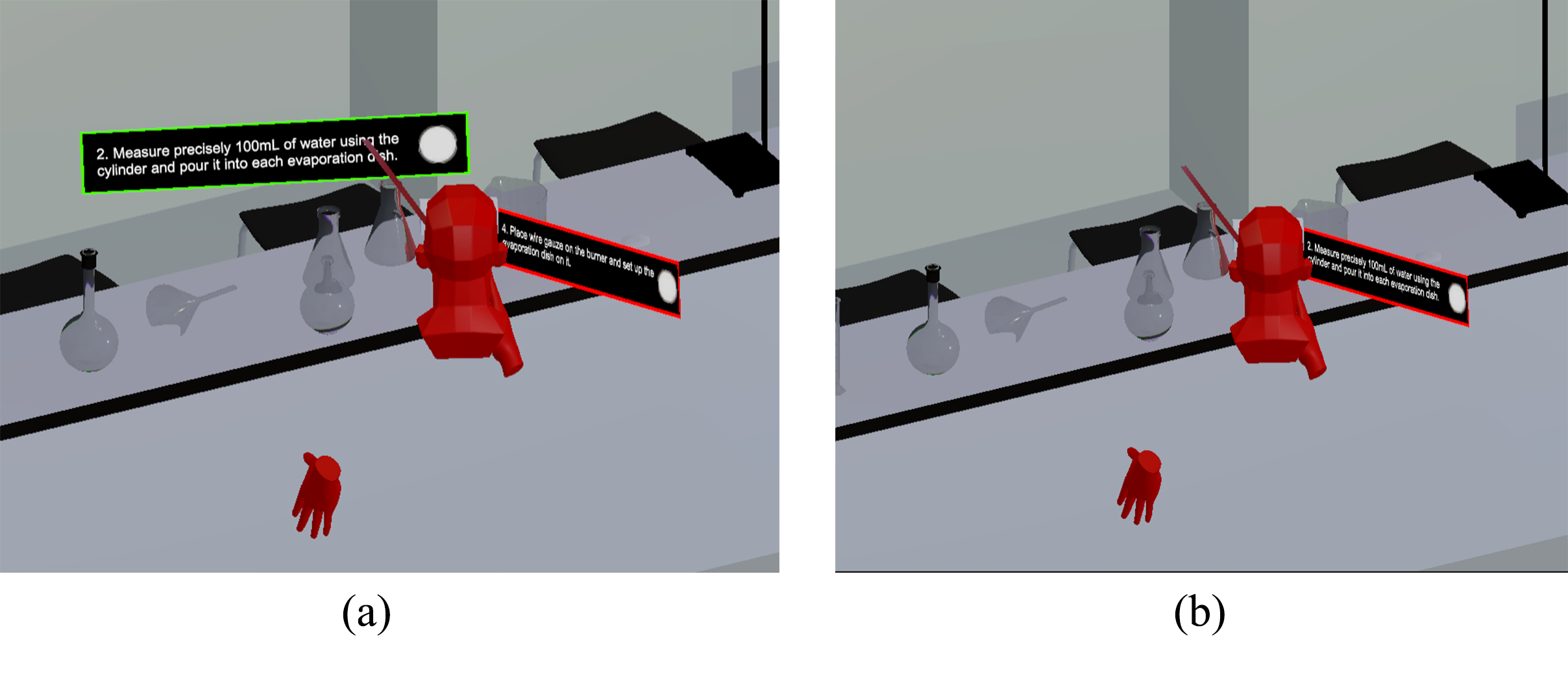}
  \vspace{-3em}
  \caption{Users can change the ownership of an incomplete DocuBit by (a) selecting them and (b) the DocuBit will be stacked to the new owner's DocuBits.}
  \label{fig:claim}
  \vspace{-1.5em}
\end{figure}

If users wish to change the ownership of a DocuBit while performing a task, they can easily modify ownership by selecting the frame of the DocuBit. Once the ownership is altered, the DocuBit seamlessly transitions into the new owner's stack of DocuBits (see Figure.~\ref{fig:claim}). Note that DocuBits that are marked ``completed'' cannot be applied for re-assigning ownership.

\section{Evaluation Study}
\label{sec:eval}
%\section{Study Design and Hypotheses}

In this section, we present the experimental design for evaluating the effectiveness of DocuBits in two distinct scenarios. First, we assess the standalone performance of DocuBits with a focus on individual learning. Second, we delve into the collaborative aspects, examining how DocuBits influence cooperative learning experiences for pairs of users engaged in a shared task with instructional documents.

For the within-subject study, participants undergo both individual and paired-user sessions. The individual study isolates DocuBits' features and examines their impact on cognitive load, immersion, learning performance, usability, and overall preference. The paired-user study extends the investigation, emphasizing collaboration-specific aspects.
The overall hypotheses (OH1, OH2, OH3, OH4, OH5) were tested across both the single-user and paired-user studies, while collaboration-specific hypotheses (CH1, and CH2) were tested with the paired-user study only. The hypotheses are rooted in the expected user experience offered by DocuBits, which aligns with the feature design rationale detailed in Section~\ref{sec:needs-finding} and Section~\ref{sec:method}.

Our hypotheses for the entire methodology were as follows:

\begin{itemize}
    \setlength\itemsep{-0.1em}
    \item \textbf{(OH1)} 
    \textit{DocuBits reduce users' cognitive load during reading-while-doing.} We anticipate that the system's ability to allow users to focus on specific task elements will result in decreased cognitive demands.
    \item \textbf{(OH2)} \textit{DocuBits enhance users' sense of immersion and presence in the virtual environment.} The tag-along-and-stick feature is expected to provide users with a more spatial sense when performing tasks.
    \item \textbf{(OH3)} \textit{Users exhibit higher learning performance with DocuBits.} The reduced cognitive load and improved spatial awareness are anticipated to contribute to superior learning outcomes.
    %\item \textbf{(OH4)} \textit{DocuBits are perceived as more usable than reading a monolithic document.} The interactive and adaptable nature of DocuBits is expected to enhance usability.
    %\item \textbf{(OH5)} \textit{Users prefer DocuBits over reading a monolithic document for instructions.} We anticipate a preference for the more dynamic and user-friendly DocuBits approach.
\end{itemize}

In the collaborative context, we introduce collaboration-specific hypotheses for the paired-user study:

\begin{itemize}
    \setlength\itemsep{-0.1em}
    \item \textbf{(CH1)} \textit{Balanced collaboration and teamwork will be enhanced with DocuBits.} DocuBits' capability to share each other's planned, ongoing, and completed/incomplete tasks will encourage equal task distribution and collaborative engagement.
    \item \textbf{(CH2)} \textit{DocuBits foster higher co-presence and collaboration levels between paired users.} The ability to see each other's actions and actively solve task distribution issues is expected to enhance collaboration.
\end{itemize}

%Validation of the overall hypotheses (OH) will be conducted across both the single user and paired user studies, while collaboration-specific hypotheses (CH) will specifically apply to the paired user study.

\subsection{Implementation Setup}
The experiment was set in a virtual chemistry lab environment built with Unity game engine. Each participant was equipped with an Oculus Quest 2 headset connected to a computer. The computer setup for each workstation included an Intel Xeon CPU and an NVIDIA GTX 1080 Ti graphics card. Participants comfortably sat on swivel chairs during the study.
For avatar movement within the virtual environment, participants had the option to turn around while seated to adjust their viewpoint. They also used controllers for aim-and-teleport actions to navigate the virtual space effectively.
In the case of the paired-user study, we ensured that both participants were physically located in the same room. This arrangement prevented audio lag issues resulting from network latency, enabling seamless communication between participants.

Regarding text presentation, we adhered to established guidelines from prior work~\cite{dingler2018vr}, employing a white Sans Serif Arial font for text display on a black background. This choice aimed to optimize text legibility and readability during the study. Text Mesh Pro was used for Unity implementation with a font size of 17.87 dmm based on previous work~\cite{kojic2020user}.

\subsection{DocuBits for Single User}
\label{sec:singleuserstudy}
The primary objective of the first study is to evaluate the efficacy of DocuBits features when utilized by a single user, specifically focusing on their impact on individual learning performance and the overall VR experience during reading-while-doing tasks.

\subsubsection{Participants}
Eighteen participants (ten females, eight males) were recruited from a convenience sample of university students for the evaluation experiment (age range $22$-$34$, $\mu=27.78$, $\sigma=4.88$). Ten participants wore glasses, one wore contact lenses, and the rest did not require vision correction. All our participants were proficient in English. Thirteen of the participants had previously experienced VR systems. 

\begin{figure}
    \centering
  \includegraphics[width=0.3\textwidth]{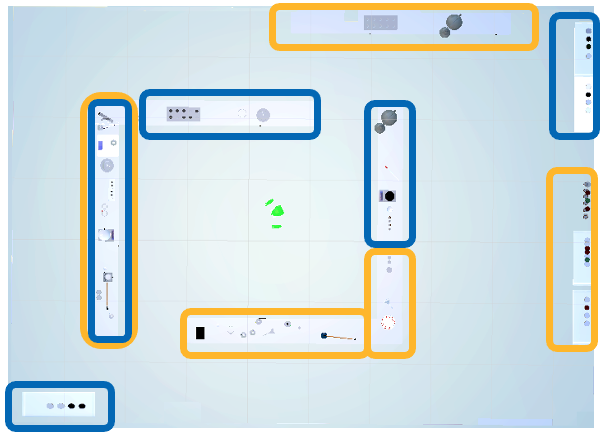}
  \vspace{-1em}
  \caption{
  A top-down view of the virtual chemistry lab, with task locations color-coded: Lab 1 in yellow and Lab 2 in blue.
  }
  \label{fig:lab_env}
  \vspace{-2em}
\end{figure}

\begin{table}[h]
    \centering
    %\small % This command sets the font size to small
    \setlength{\tabcolsep}{4pt} % Adjusts the space between columns
    \renewcommand{\arraystretch}{0.85} % Adjusts the space between rows
    \begin{tabularx}{0.4\textwidth}{X} % Use 'tabularx' and specify the table width as the text width
    \toprule[1.5pt]
    \textbf{Study Procedure Steps} \\
    \midrule[1pt]
    Instructions and Informed Consent \\
    Adjusting to the VR Headset and Environment  \\
    Training with Methods \\
    \addlinespace[1pt] % Reduced space before the double line for clarity
    \midrule[1pt]
    \textbf{Experiment 1} \\
    Lab 1 with Method A \\
    Quiz Assessment \\
    Completion of Questionnaires \\
    Lab 1 Re-performance Assessment \\
    \addlinespace[1pt] % Reduced space
    \midrule[1pt]
    \textbf{Experiment 2} \\
    Lab 2 with Method B \\
    Quiz Assessment \\
    Completion of Questionnaires \\
    Lab 2 Re-performance Assessment \\
    \bottomrule[1.5pt]
    Post-Experiment Interview \\
    \bottomrule[1.5pt]
    \end{tabularx}
    \caption{Study procedure followed by both Single-User and Paired-User studies, with methods presented in a counter-balanced order.}
    \label{tab:study_procedure}
    \vspace{-2em}
\end{table}

\begin{table*}[h]
    %\small
    \centering
    \begin{tabular}{>{\centering\arraybackslash}m{3.0cm} |>
    {\centering\arraybackslash}m{12cm}}
    \specialrule{1.5pt}{0pt}{0pt}
    WA Elements & Questions \\
    \hline\hline
      \emph{Who-Presence} &  I was aware of my collaborator's presence in the virtual environment.\\
      \emph{Who-Authorship} & I was aware of what task my collaborator was responsible for.\\
      \emph{What-Action} & I was aware of what my collaborator was doing.\\
      \emph{Where-Location} & I was aware of where my collaborator was located.\\
      \specialrule{1.5pt}{0pt}{1pt}
    \end{tabular}
    \caption{Collaboration questions based on workspace awareness from~\cite{gutwin2002descriptive}, rated on a 5-point Likert scale.}
    \label{table:collab_questions}
    \vspace{-2em}
\end{table*}
\subsubsection{Procedure}
Participants underwent a consent process, received training on the DocuBits method and a baseline monolithic document method, and completed two experiments with randomized assignments of these methods. In each experiment, participants first carried out a lab task using one of the document methods, followed by completing questionnaires about the experience. They then re-performed the same lab without any document method to assess learning retention. This process was repeated for a second lab task using the alternate document method.

%Upon arrival, participants provided demographic information, acclimatized to the VR chemistry lab via Oculus Quest 2, and received 10 minutes of training with the baseline and DocuBits. The study involved two fictional lab tasks inspired by distillation and crystallization, each comprising six interactive tasks within the VR space. Task sequences and locations were illustrated in Figure ??, with task completion times recorded from start to verbal confirmation of completion.

Upon participant arrival, we detailed the study's protocol, secured consent via signed forms, and collected demographic data. Seated in swivel chairs, participants were acquainted with the Oculus Quest 2 and given time to familiarize themselves with the VR-rendered chemistry lab. Subsequently, they were given 10 minutes of training time to familiarize themselves with the baseline and DocuBits.

The study comprised two experiments inspired by distillation and crystallization labs.  Each ``lab" involves six tasks, such as retrieving and interacting with items, which required participants to move within the VR space. 
The tasks were fictional and designed not to benefit from prior chemistry knowledge which was clearly communicated to the participants.
Task locations for each lab are illustrated in Figure~\ref{fig:lab_env}.
%Participants used the DocBits method for one of the labs and the monolithic document for the other lab.  
The order of labs was fixed, but the method the participants used (DocuBits or baseline) for each lab was randomized using a Latin Square design.
We recorded from the start of the lab until verbal confirmation of completion.

Participants were directed to follow the provided document instructions, understanding that a post-lab quiz with seven questions would assess their task comprehension. The quiz focused on the sequence of tasks, observed results, and equipment locations. After completing the quiz, they filled out questionnaires on, in this order, presence (PQ~\cite{witmer1998measuring}), immersion (IEQ~\cite{jennett2008measuring}), cognitive load (NASA-RTLX~\cite{hart2006nasa}), and system usability (SUS~\cite{bangor2008empirical}), and rated their preferences on a 5-point Likert scale. The presence questionnaire assesses the user's sense of `being there' in the virtual space, while immersion evaluates engagement with the lab experiment. This approach aligns with prior works on collaborative learning in VR employing this questionnaire~\cite{drey2022towards}.

Participants then re-performed the lab experiment without document assistance to evaluate memory-based learning performance, with metrics including time and task accuracy. Following each experiment's completion, they repeated the process with the alternate document method. After both experiments, semi-structured interviews were conducted to collect detailed feedback. The entire procedure, lasting about 90 minutes, was video-recorded for analysis. Details are in Table~\ref{tab:study_procedure}.

\subsubsection{Results}
\label{sec:single-results}
For analysis, we conducted a Wilcoxon-signed rank test for each measurement to determine the significance of our statistical results.
\begin{itemize}
\itemsep-0.1em  
    \item \textbf{Completion Time:} DocuBits allowed participants to complete the tasks slightly faster than the baseline ($Md = 11.44 min$, $IQR = 6.18$), baseline ($Md = 12.24 min$, $IQR = 8.03$). However, there were no significant difference between the two methods ($Z = -0.806, p = 0.42$). 
    
    \item \textbf{Cognitive Load:} DocuBits significantly reduced users' perceived cognitive load compared to the baseline in the single-user study. Users reported a lower cognitive load with DocuBits ($Md = 35.06$, $IQR = 22.05$) compared to the baseline ($Md = 54.62$, $IQR = 26.03$). This difference was statistically significant ($Z = -3.432, {\bf p < 0.001}$), highlighting the effectiveness of DocuBits in alleviating individual cognitive demands. This fulfills OH1:\textit{DocuBits reduce users' cognitive load during reading-while-doing.}

    \item \textbf{Immersion and Presence:} DocuBits facilitated a higher sense of immersion ($Md = 4.13$, $IQR = 0.90$) compared to the baseline ($Md = 4.22$, $IQR = 1.41$), as well as presence ($Md = 3.53$, $IQR = 1.56$, baseline: $Md = 3.24$, $IQR = 1.27$). However, no significant difference was revealed  ($Z = -0.816, p = 0.45$). While the scores are higher, it is hard to conclude that OH2:\textit{DocuBits enhance users' sense of immersion and presence in the virtual environment} is true.

    \item \textbf{Learning Performance:} In terms of learning performance, there were no significant differences in quiz results in any of the two chemistry experiment's quiz between DocuBits (quiz1: $Md = 4.00$, $IQR = 3.00$, quiz2: $Md = 4.00$, $IQR = 2.25$) and the baseline (quiz1: $Md = 3.5$, $IQR = 2.25$, quiz2: $Md = 4.00$, $IQR = 2.25$) according to the Wilcoxon signed-rank test (quiz1: $p = 0.1$, quiz2: $p = 0.32$). On the other hand, for the re-performance of tasks without instructions, users had significantly higher scores (correct steps) with DocuBits($Md = 5.00$, $IQR = 3.00$), than the baseline ($Md = 3.00$, $IQR = 2.25$) with ($Z = -3.345, {\bf p < 0.001}$). The re-performance time was also significantly faster with DocuBits ($Md = 10.51 min$, $IQR = 7.19$, baseline: $Md = 11.26 min$, $IQR = 8.56$) ($Z = -2.069,
    %{\bf p = 0.039}$). 
    {\bf p < 0.05}$). 
    This indicates that DocuBits may have helped users learn the task better, aligning with OH3:\textit{Users exhibit higher learning performance with DocuBits}.

    \item\textbf{Usability:} The SUS scores revealed that participants considered DocuBits($Md = 70.03$, $IQR = 23.49$) more usable than the baseline ($Md = 52.33$, $IQR = 25.52$). A statistical significance was revealed($Z = -3.432, {\bf p < 0.001}$).

    \item\textbf{Preference:} Participants reported a significantly higher preference towards DocuBits ($Md = 4.00$, $IQR = 2.00$) compared to the baseline ($Md = 2.50$, $IQR = 1.00$, $Z = -3.342 $ , ${\bf p < 0.001} $). 
\end{itemize}

\subsection{DocuBits for Paired Users}
In this experiment, we designed the study to require two users to work together to complete a task. Our focus was to determine the impact of having an interaction framework that allows shareable, trackable, interactive document fragments between collaborators.

\subsubsection{Participants}
We recruited 14 pairs ($N = 28$, 18 male, 10 female) from a convenience sample of university students (age range $22$-$34$, $\mu=25.7$, $\sigma=3.50$). Among the participants, 21 had prior experience with VR systems, and all were proficient in English. Participant pairs were selected such that they were already acquainted with each other, ensuring a minimum level of communication confusion due to social unfamiliarity. No participants were re-recruited from Section~\ref{sec:singleuserstudy}.

\subsubsection{Procedure}

The paired-user study also follows the procedure listed in Table~\ref{tab:study_procedure}. Here, participants were seated in swivel chairs at individual workstations in the same room. They were given time to adjust themselves to the Oculus Quest 2 and the virtual chemistry lab environment, followed by a 10-minute training to familiarize themselves with both DocuBits and the baseline.

The labs, inspired by distillation and crystallization labs, comprised eight steps distinct from those in the single-user study. Pairs proceeded through each lab using one of the two methods, with method presentation order randomized via Latin Square. Participants had complete autonomy in assigning tasks within their pair, and we recorded the time taken to complete the experiment, including task distribution, culminating when they verbally confirmed task completion. To measure task distribution among participants, we counted for each participant (i) how many tasks they finished on their own, and (ii) how many tasks were done collaboratively (completed together or taken over by another participant).

Subsequently, participants individually completed a post-trial quiz comprising seven questions without discussion. Following the quiz, they completed the same set of questionnaires as in the single-user study, assessing immersion, presence, cognitive load, usability, and preference. Additionally, we inquired about their perceived social presence in collaborative learning~\cite{lin2004social}, adapting the items to our study's context, and gathered data on collaboration-specific factors based on the workspace awareness framework~\cite{gutwin2002descriptive}, detailed in Table~\ref{table:collab_questions}. Participants then engaged in a re-performance assessment, collaboratively executing the experiment steps without written instructions.

Post-experiment, we conducted semi-structured interviews with the pairs concurrently to capture comprehensive feedback and reflections. %Participants were allowed a 10-minute intermission between trials and the flexibility to pause as needed. 
The entire procedure spanned approximately 120 minutes, with all sessions video-recorded for in-depth analysis

\begin{figure*}
	\centering
	\includegraphics[width=0.9\textwidth]{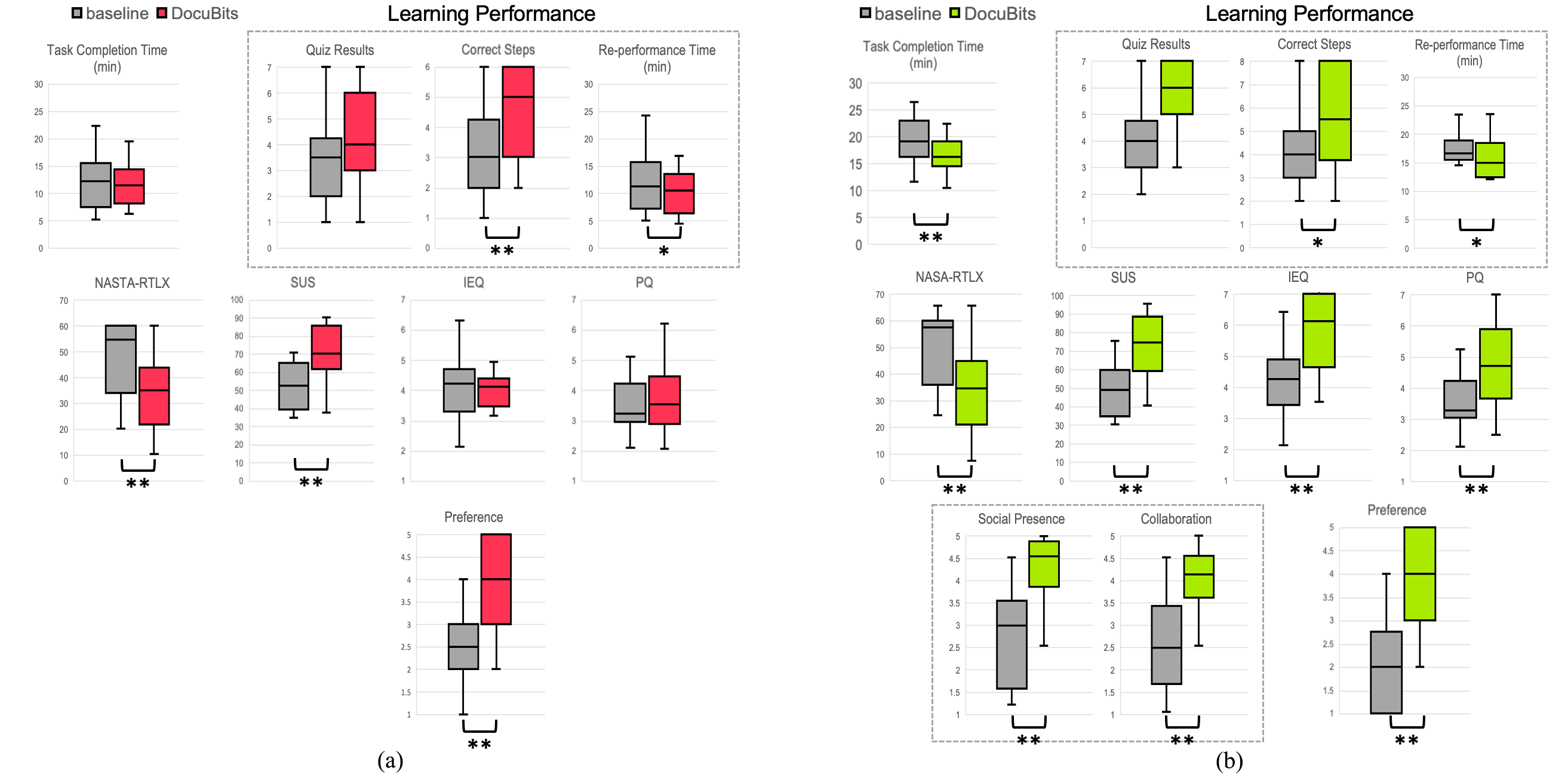}
	\caption{Statistical results for (a) the single-user study and (b) the paired-user study. In the single-user study, statistically significant results were observed for correct steps in re-performing the task, workload (NASA-RTLX), usability (SUS), and preference. In the paired-user study, statistically significant results were observed for task completion time, correct steps, re-performance time, workload (NASA-RTLX), usability (SUS), immersion (IEQ), presence (PQ), social presence, collaboration (questions described in Table.~\ref{table:collab_questions}), and preference. Asterisk (*) indicates a statistically significant difference between conditions: $p < 0.05 (*); p < 0.001 (**)$. }
    \label{fig:exp_results}
    \vspace{-2em}
\end{figure*}

\subsubsection{Results}
Similar to Section~\ref{sec:single-results}, a Wilcoxon signed-rank test was conducted for each measurement.
\begin{itemize}
\setlength\itemsep{-0.1em}
    \item \textbf{Completion Time:} DocuBits exhibited a significantly shorter completion time ($Md = 15.42 min$, $IQR = 8.81$) compared to the baseline ($Md = 17.39 min$, $IQR = 11.17$) ($Z = -3.158, {\bf p < 0.01}$).
%p = 0.002$).
    \item \textbf{Cognitive Load:} Users reported a reduced cognitive load with DocuBits ($Md = 34.64$, $IQR = 24.00$) in contrast to the baseline ($Md = 57.48$, $IQR = 24.11$). This statistically significant difference ($Z = -3.552, {\bf p < 0.001} $) underscores the effectiveness of DocuBits in alleviating user cognitive demands (OH1).

    \item \textbf{Immersion and Presence:} DocuBits facilitated a heightened sense of immersion ($Md = 6.13$, $IQR = 2.36$) compared to the baseline ($Md = 4.27$, $IQR = 1.47$) and presence ($Md = 4.72$, $IQR = 2.23$, baseline: $Md = 3.28$, $IQR = 1.18$). This significant difference ($Z = -4.213, {\bf p < 0.001} $) suggests that users felt more engaged and immersed when using DocuBits (OH2).

    \item \textbf{Task Distribution:} Task distribution was assessed by comparing the difference in the number of tasks performed by an individual within a pair. DocuBits ($Md = 1.00$, $IQR = 1.00$) demonstrated a more equal contribution among users compared to the baseline ($Md = 1.00$, $IQR = 2.00$). However, no significant difference was revealed ($Z = -1.811, p = 0.70$). On the other hand, DocuBits exhibited a higher count of tasks performed collaboratively ($Md = 2.00$, $IQR = 0.25$) compared to the baseline ($Md = 1.00$, $IQR = 2.00$). The results were statistically different ($Z = -2.739, p = 0.006$). Hence, while there are implications of equal task distribution, we can only partially support CH1: \textit{Balanced collaboration and teamwork will be enhanced with DocuBits.}.

    \item \textbf{Social Presence and Collaboration:} 
    Participants reported a higher perceived social presence with DocuBits ($Md = 4.55$, $IQR = 1.02$) compared to the baseline ($Md = 3.00$, $IQR = 1.97$). DocuBits excelled in promoting a higher level of collaboration among users ($Md = 4.14$, $IQR = 0.93$) compared to the baseline ($Md = 2.49$, $IQR = 1.76$). The observed difference is statistically significant ($Z = -4.463, {\bf p < 0.001} $), indicating a strong preference for collaborative tasks using DocuBits. This supports CH2: \textit{DocuBits foster higher co-presence and collaboration between paired users}.

    \item \textbf{Learning Performance:} In line with the single-user study, the two quiz results were comparable between DocuBits (quiz1: $Md = 6.00$, $IQR = 2.00$, quiz2: $Md = 5.00$, $IQR = 1.75$) and the baseline (quiz1: $Md = 4.00$, $IQR = 1.75$, quiz2: $Md = 6.00$, $IQR = 2.00$) (quiz1: $p = 0.46$, quiz2: $p= 0.1$). 
    However, users took less time to recreate the lab experiment with DocuBits ($Md = 13.43 min$, $IQR = 6.71$) than with the baseline ($Md = 16.02 min$, $IQR = 6.59$). This time-saving benefit with DocuBits was statistically significant ($Z = -2.417, {\bf p<0.05}$).
    %p = 0.016$ ). 
    Additionally, users demonstrated a significant difference in the number of correct steps, with DocuBits scoring higher ($Md = 5.52$, $IQR = 4.00$) than the baseline ($Md = 4.50$, $IQR = 3.31$, $Z = -2.622, {\bf p<0.01}$). 
    %p = 0.009$). 
    This supports the fact that DocuBits elicit users to have higher learning performances(OH3).

    \item \textbf{Usability:} Participants rated DocuBits higher in usability ($Md = 74.67$, $IQR = 29.26$) compared to the baseline ($Md = 49.11$, $IQR = 24.95$). This significant difference ($Z = -4.031, {\bf p < 0.001} $) affirms the enhanced usability of DocuBits in virtual collaborative environments. 

    \item \textbf{Preference:}DocuBits had a higher preference score ($Md = 4.00$, $IQR = 2.00$) than the baseline ($Md = 2.00$, $IQR = 1.75$), which was statistically significant ($Z = -4.266,{\bf p < 0.001} $).
\end{itemize}
Here, our approach, DocuBits, consistently outperformed the baseline across all metrics, establishing its effectiveness in collaborative learning scenarios.

\section{Discussion}
In this section, we discuss the results from the two evaluation studies presented in Section~\ref{sec:eval}. We focus on how the results support or do not support our hypotheses, using quotes from participants from the post-experiment interviews for illustration. Participants are numbered continuously, with P1-P18 from the single-user study and P19-P48 from the paired-user study.

\subsection{DocuBits for Facilitating Collaboration}
Our result analysis revealed DocuBits offered beneficial results in both the single-user study (OH1, OH3) and the paired-user study (OH1, OH2, OH3, CH1, CH2).
This indicates that DocuBits could be a valuable tool in either setting, with the added benefits of greater immersion and collaboration in the paired setting. Particularly, the higher number of tasks completed collaboratively by paired participants using DocuBits supports the notion of balanced collaboration (CH1). This trend demonstrates DocuBits' potential to foster an equitable task distribution among users.

Participants emphasized that the interface facilitated collaborative decision-making sessions, aided visually by DocuBits. The color-coded interface allowed for immediate tracking of the distribution of work between users. P12 highlighted, ``\textit{The process of selecting which task I am going to perform made me more conscious of how much work I was doing. I wanted to make sure my partner was not doing all the work for me.}" Even during the re-performance test, participants exhibited a better recollection of who had performed which task with DocuBits, attributing it to the specific procedural guidance provided. Similarly, P7 expressed, ``\textit{I was constantly aware of the workload balance, thanks to the visual cues from DocuBits.}" Pair P43 and P44 noted the shared features of DocuBits facilitated their cooperative work. P43 remarked, ``\textit{That floating DocuBit with an `X' on it really grabbed my attention. I just had to jump in and help my partner out so we could get it to stay put!}"

\subsection{Preferred Method for Creating DocuBits}
In the observation of participants' preferences for creating DocuBits, a prominent trend emerged, showcasing a clear inclination towards the stepped method. This preference was particularly pronounced when participants engaged in individual tasks. P19 reflected, ``\textit{The stepped method provided a structured and organized approach. It felt like a natural way to break down complex instructions into manageable steps, especially when I was working through them on my own. It helped maintain a clear sequence and aided in a more systematic execution of tasks.}"

On the other hand, some participants (P20, P21, P24, P25, P31) mentioned that their preference for the stepped method was due to convenience, but they acknowledged the potential need for the highlighting method in more complex tasks. Particularly, the user pair P24 and P25 collectively recalled, ``\textit{the stepped method was easier with the instructions we had, but in cases where a single step involves multiple substeps, the highlight method would come in handy.}'' P31 suggested the usefulness of being able to switch between methods, saying, ``\textit{Ultimately, I would prefer if I could do both combined. Sometimes I wanted to make minor adjustments to the pre-fragmented steps and I don't want to spend too much time highlighting every step.}''

This inclination towards a stepped approach underscores the importance of tailoring the DocuBits interface to individual work styles, ensuring that the method of creating these instructional cues aligns with user preferences. An intuitive and adaptable creation method could significantly contribute to the user's efficiency and comfort, ultimately enhancing the overall user experience.

\subsection{Need for Ownership Support}
Participants articulated a compelling need for enhanced ownership support, a crucial aspect when collaborating on tasks within the DocuBits framework. The majority of the participant pairs in the paired user study pointed out how they liked the idea of transferring ownership of a DocuBit. A participant pair, P19, P20 stated, ``The smooth transitions of ownership not only enhances the collaborative experience but also allowed more freedom in collaboration styles and preferences.''

A noteworthy perspective surfaced, with participants expressing a desire to assign the same task to multiple individuals and have it automatically marked as complete once any one of them finished. As P34 emphasized, ``\textit{In collaborative settings, there are often times you wish to work on the same task with your collaborator or allow whoever finishes the previous task first to take on the task. It would be more practical if there could be a way we can have access to a DocuBit at the same time. The current version only allows one owner per DocuBit.}" This participant insight underscores the significance of refining collaborative features within DocuBits to accommodate diverse task distribution scenarios.

\subsection{Observed Collaboration Styles}
Diverse collaboration styles were observed among participants, revealing intriguing insights into task allocation strategies within the DocuBits framework. Some users used a strategy of pre-placing all DocuBits before initiating the procedural task, reflecting a meticulous planning approach. P38 elucidated, ``{\em I preferred setting everything up first. It provided a comprehensive visualization of the entire process before diving into the actual tasks. This strategic planning helped me allocate tasks more effectively.}"

Additionally, users displayed a proclivity for dividing tasks evenly or allocating specific types of tasks to different users, showcasing the flexibility and adaptability of DocuBits to varied collaboration styles. For instance, participants chose to exclusively handle apparatus-related tasks as a specialized allocation strategy. These insights highlight the diverse ways users approach collaborative work within the DocuBits environment, emphasizing the need for a flexible and intuitive tool that accommodates these varied preferences.

\subsection{Effect of Texts on Collaboration}
Participants shed light on the nuanced impact of text-based instructions on collaboration, revealing an interesting dynamic within the DocuBits framework. Some participants expressed a need for integration with demonstration videos, envisioning a more comprehensive guidance approach. P44 suggested, ``\textit{Having the option to seamlessly switch between text and video instructions could significantly enhance task distribution and performance. It caters to different learning preferences and ensures a more versatile collaboration experience.}"

This participant insight unveils a potential avenue for future work, pointing towards the integration of textual and visual guidance within DocuBits. Such a feature could provide users with versatile options for task execution and distribution, further enhancing the collaborative and instructional capabilities of the tool. As DocuBits evolves, addressing these nuanced preferences for instructional content could substantially contribute to the framework's adaptability and user-centric design.

\section{Limitations and Future work}
Our method's efficacy in comparison with documents featuring scrolls and pages warrants exploration. The frequency of feature usage is anticipated to vary based on task characteristics. For instance, users might utilize the highlight feature more for excessively long steps or when tasks are perceived as intricate.

The current implementation focuses on collaboration between two users. Scaling the tool for more than two users necessitates refined design considerations. Improved interaction methods for claiming DocuBit ownership should address potential confusion regarding task attribution. Integration with the spatial map could assist in tracking task locations and current user engagement. Additionally, enhancing color-coded ownership indicators with user-specific icons can enhance clarity.

As for future work, we envision advancing DocuBits into a seamless pipeline. Future work will involve:
\vspace{-0.5em}
\begin{itemize}
    \setlength \itemsep{-0.5em}
    \item \textbf{Automatic Instructional Document Integration:} Development of an automatic pipeline to scan and import instructional documents into DocuBits.
    \item \textbf{Automated Performance Evaluation:} Implementation of an automated performance evaluation mechanism to assess task completion.
    \item \textbf{Instructor-Side Connectivity:} Sharing task performance results with the instructor side for real-time analysis of student performance and identification of areas requiring assistance.
\end{itemize}
\vspace{-0.5em}
Our current prototype requires clear numbering or manual entry of instructions into DocuBits. Future implementations aim to extend this by facilitating the easy selection of typical separators. We foresee establishing labeled ``anchor points" during authoring to automate DocuBit placement. Expansion beyond two users is anticipated, with a proposed special ``birdseye" view for instructors in classroom scenarios. Furthermore, we envision extending DocuBits to Mixed Reality, utilizing MR anchors in real space instead of VR coordinates.

\section{Conclusion}
We introduced DocuBits, a novel method for decomposing monolithic documents into portable interactive elements. Informed by insights from a needs-finding study with 7 experienced educators, we delineated key design considerations for VR interfaces handling text instructions. DocuBits facilitated procedural task completion by (i) fragmenting text instructions, (ii) anchoring them to task locations, and (iii) enabling progress monitoring and sharing among users. In a comprehensive two-fold user study—first with a single user and then with paired users—we established that DocuBits enhanced learning performance in task recall and execution, while also positively impacting perceived workload, presence, immersion, and usability.

We further discussed how DocuBits fostered collaboration, user strategies in creating DocuBits, necessary support for multi-user DocuBits assignments, observed collaboration styles between paired users, and the potential influence of text instructions on collaboration dynamics. We anticipated DocuBits' extension for supporting automated performance evaluation in VR training applications and its integration into AR/MR contexts.
\bibliographystyle{abbrv-doi}

\bibliography{docubits.bib}

\begin{thebibliography}{10}
\bibitem{gould1987reading}
J.~D. Gould, L.~Alfaro, R.~Finn, B.~Haupt, and A.~Minuto.
\newblock Reading from crt displays can be as fast as reading from paper.
\newblock {\em Human factors}, 29(5):497--517, 1987.
\end{thebibliography}
\end{document}